\documentclass[prd,reprint,amsmath,amssymb,preprintnumbers,superscriptaddress,nofootinbib]{revtex4-1}

\usepackage{subfigure}
\usepackage{enumitem}
\usepackage{color}
\usepackage{graphicx}

\def\beq{\begin{equation}}
\def\eeq{\end{equation}}
\def\bea{\begin{eqnarray}} 
\def\eea{\end{eqnarray}}
\def\nn{\nonumber}
\def\gev{\rm GeV}

\def\kev{\rm keV}
\def\ev{\rm eV}
\def\eps{\varepsilon}

\def\vpq{f_a}
\newcommand{\lsim}{
\mathrel{\hbox{\rlap{\hbox{\lower4pt\hbox{$\sim$}}}\hbox{$<$}}}}
\newcommand{\gsim}{
\mathrel{\hbox{\rlap{\hbox{\lower4pt\hbox{$\sim$}}}\hbox{$>$}}}}

\begin{document}

\preprint{CTPU-17-15}
\title{Dark photon relic dark matter production through the dark axion portal}

\author{Kunio Kaneta}
\affiliation{Center for Theoretical Physics of the Universe, Institute for Basic Science (IBS), Daejeon 34051, Korea}
\author{Hye-Sung Lee}
\affiliation{Center for Theoretical Physics of the Universe, Institute for Basic Science (IBS), Daejeon 34051, Korea}
\author{Seokhoon Yun}
\affiliation{Center for Theoretical Physics of the Universe, Institute for Basic Science (IBS), Daejeon 34051, Korea}
\affiliation{Department of Physics, KAIST, Daejeon 34141, Korea}

\date{April 2017}
\begin{abstract}
We present a new mechanism to produce the dark photon ($\gamma'$) in the early universe with a help of the axion ($a$) using a recently proposed dark axion portal.
The dark photon, a light gauge boson in the dark sector, can be a relic dark matter if its lifetime is long enough.
The main process we consider is a variant of the Primakoff process $f a \to f \gamma'$ mediated by a photon, which is possible with the axion--photon--dark photon coupling.
The axion is thermalized in the early universe because of the strong interaction and it can contribute to the non-thermal dark photon production through the dark axion portal coupling.
It provides a two-component dark matter sector, and the relic density deficit issue of the axion dark matter can be addressed by the compensation with the dark photon.
The dark photon dark matter can also address the reported 3.5 keV $X$-ray excess via the $\gamma' \to \gamma a$ decay.
\end{abstract}
\maketitle

\section{Introduction}
Conventional attitude toward the new physics is based on the presumption that the new particles have a similar coupling size as the standard model (SM) couplings.
There are many popular models leading to this including supersymmetry, extra dimension, grand unified theories.
In this approach, after one energy scale is probed at some level, it is essential to increase the energy of the experiments to find a new uncovered particle, which typically means building a larger, higher energy beam facility.
Among them are the currently running 13 TeV Large Hadron Collider and an envisioned 100 TeV collider.
This line of research is categorized as the Energy Frontier \cite{Gershtein:2013iqa}.

There has been an alternative attitude toward the new physics, which are perhaps less popular yet long-standing.
The new particles may have a significantly smaller coupling, at least to the SM particles, which makes it hard to detect in the typical experiments designed to probe particles of an ordinary size coupling.
Therefore they can be very light, and getting a higher energy may not be necessary to search for them.
It is more important to have enough statistics (and even to develop new search schemes), which is called the Intensity Frontier \cite{Essig:2013lka}.

General attitude towards the new particles also affects the Cosmic Frontier \cite{Kusenko:2013saa}, including the dark matter search.
The typical WIMP (weakly interacting massive particle) search using the nuclear recoil assumes a weak scale new particle for a dark matter  \cite{Cushman:2013zza,Tan:2016zwf,Akerib:2016vxi,Alexander:2016aln}.
Yet there are very light (say, below GeV scale) dark matter candidates which require completely different search schemes.

Axion (spin-0 light pseudoscalar) and dark photon (spin-1 light vector boson) are two popular candidates of the light, feebly interacting new particles.
Depending on their mass and coupling, each can be a dark matter candidate.
Although less popular than the WIMP dark matter candidate, each of the two has its own history of many theoretical and experimental investigations.

Recently, it was pointed out a genuinely new coupling $G_{a\gamma\gamma'}$ that combines the axion ($a$) and dark photon ($\gamma'$) is possible \cite{Kaneta:2016wvf}.
Introduction of this coupling inevitably brings $G_{a\gamma'\gamma'}$ coupling too.
They are collectively named ``dark axion portal'' \cite{Kaneta:2016wvf}.
(In some sense, the $G_{a\gamma'\gamma'}$ coupling was first studied in the mirror world models where a massless mirror photon couples to the axion \cite{Berezhiani:2000gh,Ejlli:2016asd}. See also a recent study in the cosmological relaxation mechanism for a solution to the hierarchy problem \cite{Choi:2016kke}.)
The dark axion portal couplings can be as large as the typical axion coupling $G_{a\gamma\gamma}$ or even larger depending on the model.

As both the axion and dark photon can be the dark matter candidates, the new portal is important making a connection of the two dark matter candidates.
In Ref.~\cite{Kaneta:2016wvf}, a specific model `dark KSVZ model' was presented to realize the dark axion portal and an illustration was made how the dark photon can be produced in the early universe using the $G_{a\gamma'\gamma'}$ coupling.
The dark photon dark matter produced with the help of the axion can compensate the deficit relic density which is a long-standing problem of the axion dark matter for the $f_a \gsim 10^{11} ~\gev$ where $f_a$ is the Peccei-Quinn (PQ) symmetry breaking scale.

In this paper, we mainly exploit the other coupling $G_{a\gamma\gamma'}$ and investigate a new dark photon production scenario in the early universe.
With a new coupling, a novel dark photon production channel $f a \to f \gamma'$ is possible.
It is similar to the Primakoff process using $G_{a\gamma\gamma}$ coupling, and we call it `dark Primakoff' process.
Interestingly, this coupling allows the $\gamma' \to \gamma a$ decay that can address the reported 3.5 keV $X$-ray excess \cite{Bulbul:2014sua,Boyarsky:2014jta,Riemer-Sorensen:2014yda,Jeltema:2014qfa,Boyarsky:2014ska,Iakubovskyi:2015dna}.
We will also elaborate the $G_{a\gamma'\gamma'}$ process providing more detailed description compared to the brief illustration in Ref.~\cite{Kaneta:2016wvf}.

The rest of this paper is organized as followings.
In Secs.~\ref{sec:PQandAxion} and \ref{sec:DarkGaugeBoson}, we give brief overviews on the axion  and dark photon physics, respectively.
In Sec.~\ref{sec:DarkAxionPortal}, we discuss the dark axion portal and benchmark points we want to study in this paper.
In Sec.~\ref{sec:DarkPhotonProduction}, we investigate the dark photon production in the early universe using the dark axion portal couplings $G_{a\gamma\gamma'}$ and $G_{a\gamma'\gamma'}$.
In Sec.~\ref{sec:3.5keV}, we address the 3.5 keV $X$-ray excess from the dark photon decay.
We devote Sec.~\ref{sec:discussions} to discussions on some issues.
We summarize our results in Sec.~\ref{sec:summary}.

\section{Overview of Peccei-Quinn symmetry and axion models}
\label{sec:PQandAxion}
The strong $CP$ problem is one of the long-standing issues in particle physics.
Once we introduce the vacuum angle $\bar \theta_s$ as $\bar \theta_s G_{\mu\nu} \tilde G^{\mu\nu}$, we encounter the $CP$ violation in QCD.
From the experimental side, the measurement of the neutron dipole moment gives a stringent constraint, $\bar\theta_s\lesssim 10^{-10}$ \cite{Baker:2006ts}, while from the theoretical side, no reason exists to keep its value that small as $\bar\theta_s\sim {\cal O}(1)$ is naturally expected.
One of the promising solutions for the strong $CP$ problem is the global PQ symmetry \cite{Peccei:1977hh,Peccei:1977ur} of which breaking gives rise to the QCD axion, a pseudo-Nambu-Goldstone boson, that makes the $\bar\theta_s$ vanish dynamically.

In the original axion models \cite{Peccei:1977hh,Peccei:1977ur,Weinberg:1977ma,Wilczek:1977pj}, the $U(1)_{\rm PQ}$ symmetry is supposed to be spontaneously broken down at the electroweak scale, $v_{\rm EW}$, and the massless axion emerges.
The non-perturbative QCD effect explicitly violates the PQ symmetry, and the axion potential is lifted up, allowing the axion to acquire finite mass of the order of $\Lambda_{\rm QCD}^2/v_{\rm EW}$ with $\Lambda_{\rm QCD}$ being the confinement scale.
On the other hand, due to the relatively large coupling among the axion and the SM particles, this ${\cal O} (100 ~\kev)$ axion model has been excluded by the rare decay measurements of mesons \cite{Bardeen:1986yb}.

The invisible axion models were then proposed to evade various experimental constrains.
The Kim-Shifman-Vainshtein-Zakharov (KSVZ) \cite{Kim:1979if,Shifman:1979if} and Dine-Fischler-Srednicki-Zhitnitsky (DFSZ) \cite{Dine:1981rt} axion models are known as viable realizations.
Their main idea is to raise the breaking scale of the $U(1)_{\rm PQ}$, much larger than the electroweak scale, $f_a\gg v_{\rm EW}$, by introducing new particles, which make the axion interactions feeble as they are proportional to $1/f_a$.

Although the invisible axion still suffers from astrophysical constraints with $f_a \lsim 10^9~\gev$, it has a great merit for dark matter physics, i.e. the coherent oscillation of the axion can constitute the dark matter in the universe.
When the PQ symmetry breaking takes place in the early universe, the axion field value is randomly distributed along the degenerated vacuum with the angle $\theta_i=a_{\rm ini}/f_a$ in which $a_{\rm ini}$ is the initial field value of the axion.
After the QCD phase transition occurs, the axion starts to oscillate at the time when the cosmic expansion becomes slow compared to the oscillation frequency, and thus the energy density of this oscillation plays a role of the cold dark matter (CDM) ~\cite{Preskill:1982cy,Abbott:1982af,Dine:1982ah}.
This scenario is called the misalignment mechanism, and the resultant CDM abundance is given by
\bea
\Omega_{a}h^2 \simeq 0.12 \times \Big(\frac{\vpq}{5.4 \times 10^{11} ~\gev}\Big)^{1.19} \, \theta_i^2\,  F(\theta_i)
\eea
where the anharmonic effect in the axion potential is taken into account by the $F(\theta_i)$ \cite{Bae:2008ue}.
It should be noted that for $\theta_i \lesssim 1$ the anharmonic effect is negligible, and we can take $F(\theta_i)\sim1$, otherwise $F(\theta_i)$ gets monotonically increasing up to a few factor.
To evade the overproduction of the axion CDM, we need $f_a \lsim 10^{12} ~\gev$ for $\theta_i={\cal O}(1)$.

\section{Overview of Vector portal and Dark photon}
\label{sec:DarkGaugeBoson}
A gauge boson much lighter than the electroweak scale can be constructed in various scenarios \cite{ArkaniHamed:2008qp,Cheung:2009qd,Lee:2016ejx}.
It has decades of history (for instance, see Ref.~\cite{Fayet:1980rr}) with different names.
A light gauge boson physics has motivations from the dark matter related phenomena (such as the explanation of the positron excess \cite{ArkaniHamed:2008qn}, self-interacting dark matter \cite{Tulin:2012wi}) as well as non-dark matter related  phenomena (such as the muon anomalous magnetic moment anomaly \cite{Gninenko:2001hx,Fayet:2007ua,Pospelov:2008zw}).

For such a light gauge boson to survive all the experimental constraints, it should have a very small coupling to the SM fermions.
Typically a dark gauge symmetry $U(1)_\text{Dark}$ is assumed under which the SM fermions do not carry a charge, and its gauge boson couples to the SM fermions only through a small mixing with the SM gauge boson \cite{Holdom:1985ag}.
The kinetic mixing of the $U(1)_\text{Dark}$ with the SM $U(1)_Y$ is described by the parameter $\eps$.
\bea
{\cal L}_\text{kinetic} = - \frac{1}{4} B_{\mu\nu} B^{\mu\nu} + \frac{\eps}{2 \cos\theta_W} B_{\mu\nu} Z'^{\mu\nu} - \frac{1}{4} Z'_{\mu\nu} Z'^{\mu\nu}  \nn \\
\eea

For a light $Z'$, the interaction lagrangian of the physical $Z'$ is given by \cite{darkPhotonCoupling}
\bea
{\cal L}_\text{int} \simeq - \eps e J_\text{EM}^\mu Z'_\mu - \eps \tan\theta_W \frac{m_{Z'}^2}{m_Z^2} g_Z J_\text{NC}^\mu Z'_\mu \label{eq:Lint}
\eea
where $e$ and $g_Z$ are the electromagnetic coupling and weak neutral current coupling ($g_Z = g / \cos\theta_W$), respectively.
$J_\text{EM}$ ($J_\text{NC}$) is the electromagnetic (weak neutral) current.
Eq.~\eqref{eq:Lint} suggests that we can ignore its coupling to the weak neutral current as long as the ratio $m_{Z'} / m_Z$ is sufficiently small, and this is the limit we take in this paper.
As it couples predominantly to the electromagnetic current, it is typically called the dark photon.

We note that the specific form of the coupling to the weak neutral current in Eq.~\eqref{eq:Lint} depends on the mechanism how the $Z'$ gets a mass.
For instance, if the Higgs sector is based on the two Higgs doublet model, a light $Z'$ may still have a sizable coupling to the weak neutral current as shown in the dark $Z$ model \cite{Davoudiasl:2012ag}.
(In these models, a charged Higgs is introduced whose major decay mode can be strikingly different due to the light gauge boson \cite{Lee:2013fda,Davoudiasl:2014mqa,Kong:2014jwa,Kim:2014ana}.)
We will consider more general cases including the case the neutral current contribution is important in the future work.

\section{Dark axion portal}
\label{sec:DarkAxionPortal}
Now let us consider the framework in which the axion and dark photon co-exist.
The axion portal to the SM gauge fields is given by
\bea
{\cal L}_\text{axion portal} = \frac{G_{agg}}{4} a G_{\mu\nu}\tilde G^{\mu\nu} + \frac{G_{a\gamma\gamma}}{4} a F_{\mu\nu}\tilde F^{\mu\nu} + \cdots ~~
\eea
where $G_{\mu\nu}$ and $F_{\mu\nu}$ are the field strength of the gluon and photon, respectively, and  the tilde represents the dual of the field strength.
In addition, a new portal coupling, the dark axion portal \cite{Kaneta:2016wvf}, can emerge by introducing the dark photon, which is given by
\bea
{\cal L}_\text{dark axion portal} = \frac{G_{a\gamma^\prime\gamma^\prime}}{4}  a Z'_{\mu\nu}\tilde Z'^{\mu\nu}
+\frac{G_{a\gamma\gamma^\prime}}{2} a F_{\mu\nu}\tilde Z'^{\mu\nu} ~~~
\eea
where $F_{\mu\nu}$ ($Z'_{\mu\nu}$) is the field strength of the photon (dark photon).
Hereafter we denote the dark photon as $\gamma'$ in the basis obtained by eliminating the kinetic mixing.

\begin{table}[bt]
\begin{tabular}{c|ccc|cc}
\hline
~Field~ & $SU(3)_C$ & $SU(2)_L$ & $U(1)_Y$ & $U(1)_\text{Dark}$ & $U(1)_{\rm PQ}$ \\
\hline
$\psi$    &  $3$ & $1$ & $Q_\psi$  & $D_\psi$   & $PQ_\psi$ \\
$\psi^c$ & $\bar 3$ & $1$ & $-Q_\psi$ & $-D_{\psi}$ & $PQ_{\psi^c}$ \\
\hline
$\Phi_{PQ}$    & $1$ & $1$ & $0$ & $0$ & $PQ_\Phi$ \\
$\Phi_D$    & $1$ & $1$ & $0$ & $D_\Phi$ & $0$ \\
\hline
\end{tabular}
\caption{New fields and their charge assignments in the dark KSVZ model.
$Q_\psi$ ($D_\psi$) is the electromagnetic (dark) charge of the exotic colored fermion $\psi$.}
\label{tab:fields}
\end{table}

\begin{table}[b]
\begin{tabular}{c|cccc}
\hline
Case ~&~ $G_{agg}$ & $G_{a\gamma\gamma}$ & $G_{a\gamma\gamma'}$ & $G_{a\gamma'\gamma'}$ \\
\hline
(i) $Q_\psi = 0$, $D_\psi = 3$ ~&~ $g_S^2$ & $0 $ & $0$  & $e'^2 (54)$ \\
(ii) $Q_\psi = -\frac{1}{3}$, $D_\psi = 3$ ~&~ $g_S^2$ & $e^2 (2/3)$ & $ee' (-6)$ & $e'^2 (54)$ \\
\hline
\end{tabular}
\caption{The relevant axion portal couplings and dark axion portal couplings. For all terms, a common factor $\frac{1}{8\pi^2} \frac{PQ_\Phi}{f_a}$ is omitted.}
\label{tab:cases}
\end{table}

A simple realization of the dark axion portal is the dark KSVZ model considered in Ref.~\cite{Kaneta:2016wvf}.
The new fields and their charges in this model are described in Table~\ref{tab:fields}, where $\psi$ and $\psi^c$ are introduced as vector-like colored fermions, and $\Phi_{PQ}$ and $\Phi_D$ are singlet scalar fields which spontaneously break the $U(1)_{\rm PQ}$ and $U(1)_{\rm Dark}$ by developing non-zero vacuum expectation value, respectively.
Throughout this paper, we discuss dark matter production and relevant phenomenology by taking this setup as an example.
Above the QCD scale, both the axion and dark axion portals are given by
\bea
G_{agg} &=& \frac{g_S^2}{8\pi^2} \frac{PQ_\Phi}{f_a}, \\
G_{a\gamma\gamma} &=& \frac{e^2}{8\pi^2} \frac{PQ_\Phi}{f_a} \big[ 2 N_C Q_\psi^2 \big], \\
G_{a\gamma\gamma'} &=& \frac{e e'}{8\pi^2} \frac{PQ_\Phi}{f_a} \big[ 2 N_C D_\psi Q_\psi \big] + \eps G_{a\gamma\gamma}, \label{eq:GaAAprime} \\
G_{a\gamma'\gamma'} &=& \frac{e'^2}{8\pi^2} \frac{PQ_\Phi}{f_a} \big[ 2 N_C D_\psi^2 \big] + 2 \eps G_{a\gamma\gamma'}, \label{eq:GaAprimeAprime}
\eea
at the leading order with respect to $\eps$, where $N_C = 3$ is the color factor, $g_S$ is the $SU(3)_C$ gauge coupling, and $e'$ is the $U(1)_\text{Dark}$ gauge coupling.
Here, we define $f_a^2 = 2 PQ_\Phi^2 \langle \Phi_{PQ} \rangle^2 $ and in the following discussion we will take $PQ_\Phi = -(PQ_\psi+PQ_{\psi^c}) = 1$ for the illustration purpose.

\begin{figure}[t]
\centering
\includegraphics[width=0.25\textwidth]{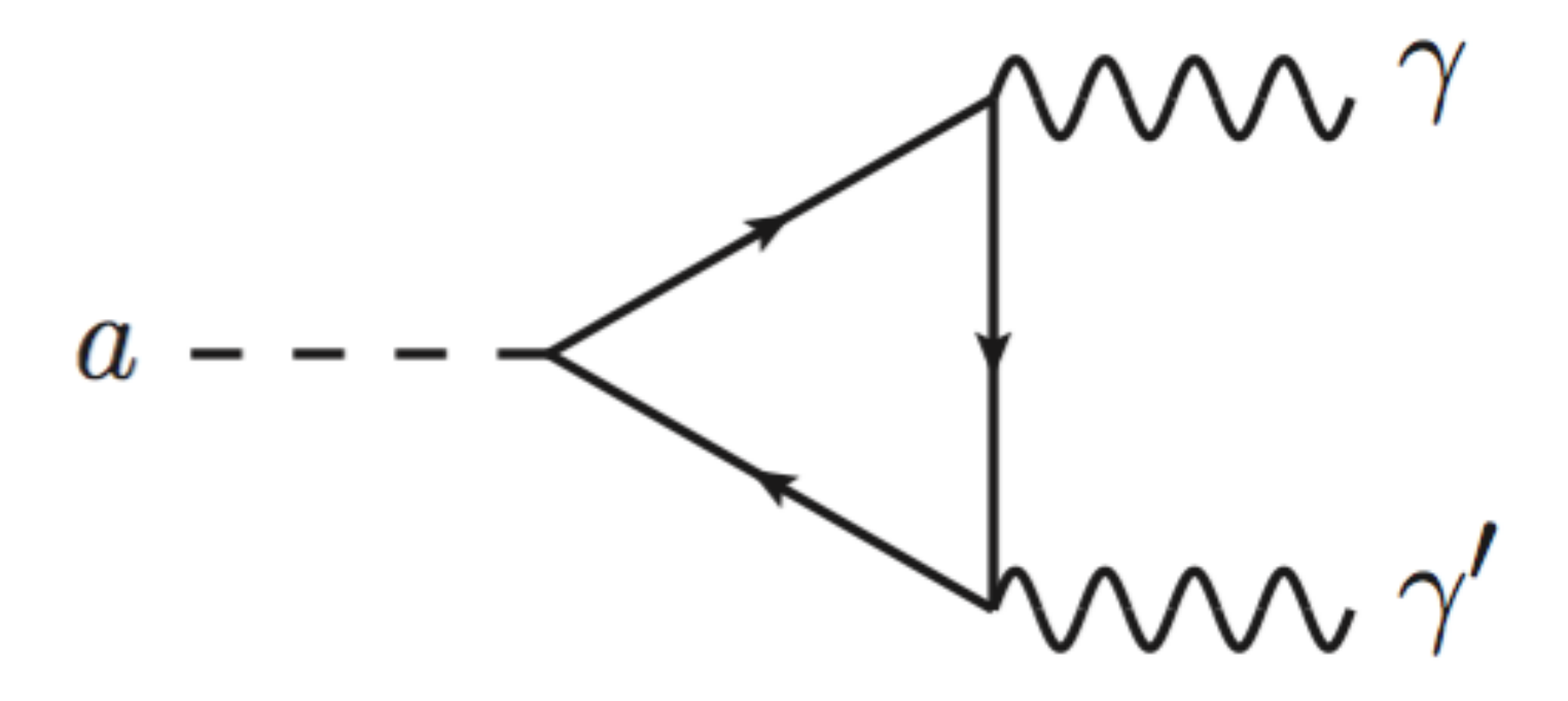}
\hfill
\caption{Dark Primakoff mechanism is delivered by the dark axion portal coupling $G_{a\gamma\gamma'}$.
The fermions inside the loop need to be charged under all $U(1)_{\rm PQ}$, $U(1)_\text{EM}$, and $U(1)_\text{Dark}$.}
\label{fig:triangle}
\end{figure}

We emphasize that the dark axion portal (in other words, vector-axion portal) is not a product of two other portals (vector portal and axion portal).
The second terms in Eqs.~\eqref{eq:GaAAprime} and \eqref{eq:GaAprimeAprime} are from that product, but the first terms are not.
The first terms originate from the exotic fermions in the triangle loop that couple to the axion, photon, dark photon directly (see Fig.~\ref{fig:triangle}).

In the next section, we will study the following two cases as the benchmark scenarios,
\begin{itemize}[noitemsep]
\item[] Case (i) :  $Q_\psi = 0$ and $D_\psi = 3$ , \\
\item[] Case (ii) : $Q_\psi = -1/3$ and $D_\psi = 3$ .
\end{itemize}
For definiteness we will assume $\eps \simeq 0$.
The axion and dark axion portal couplings in these cases are given in Table~\ref{tab:cases}.

Before closing this section, let us comment on taking the vanishing kinetic mixing.
This is possible because the $\varepsilon$ is a free parameter at the tree level.
On the other hand, this parameter choice does not hold if there is a radiatively induced kinetic mixing.
For instance, in the case (ii), since the exotic fermion is charged under both $U(1)_{\rm EM}$ and $U(1)_{\rm Dark}$, the mixing between $\gamma$ and $\gamma'$ is induced at one loop level.
The order of magnitude of the induced kinetic mixing is estimated by following the renormalization group (RG) evolution, where we define $\beta_\varepsilon \equiv d\varepsilon / d\log\mu$.
For the RG scale $\mu$ above the exotic fermion mass $m_\psi$, we have
\bea
\beta_{\varepsilon} \left(\mu > m_\psi\right) = \frac{ee'}{6\pi^2}N_{C,\psi} Q_\psi D_\psi,
\eea
where $N_{C,\psi} = 3$ is the number of color degrees of freedom.
For instance, if we take $\varepsilon=0$ at a certain scale $\Lambda$ higher than $m_\psi$, such as the grand unification scale, we obtain the induced value of the $\varepsilon$
\bea
\eps_\text{induced} = \frac{ee'}{6\pi^2} N_{C,\psi}Q_\psi D_\psi \log\big(\frac{m_\psi}{\Lambda}\big)
\eea
at an energy scale lower than $m_\psi$.
It should be noted that for $\mu<m_\psi$, the RG running of $\varepsilon$ is given by
\bea
\beta_{\varepsilon} \left(\mu < m_\psi\right) = \varepsilon \frac{e^2}{6\pi^2}\sum_f N_{C,f}Q_f^2,
\label{eq:betalow}
\eea
where $Q_f$ and $N_{C,f}$ are the electric charge and and color factor of the SM fermion $f$.
Since there is an additional $\varepsilon$ in Eq.~(\ref{eq:betalow}), the RG running of $\varepsilon$ below $m_\psi$ is negligibly small.

Therefore, the radiatively induced kinetic mixing is estimated as $\varepsilon_\text{induced} \approx 0.015 \, e' \log(m_\psi / \Lambda)$ for $Q_\psi D_\psi = 1$, and $\varepsilon_\text{induced} \sim - \, {\cal O} (10^{-2})$ if we take $e' = 0.1$ with $\Lambda \sim 10^{16} ~\gev$ (typical GUT scale) and $m_\psi \sim f_a $ ($10^9 - 10^{12} ~\gev$).
The induced value itself is inconsistent with astrophysical observations and the beam dump experiments for the keV-MeV scale dark photon \cite{Essig:2013lka}.

On the other hand, the $\eps$ value at the UV scale ($\Lambda$) is not determined in general, and we can take $\eps = \eps(\Lambda) + \eps_\text{induced}$ sufficiently small at the cost of fine-tuning.
Alternatively, it is also possible to suppress the radiatively induced $\varepsilon$ by introducing another exotic fermion having the same mass as $\psi$ and opposite dark charge, by which the cancellation between the contributions from these two fermions can occur \cite{Holdom:1985ag,Davoudiasl:2012ag}.
Since our discussion is independent from the number of exotic fermions, our main result does not change as long as additional fermions do not contribute to the dark axion portal significantly.

\section{Dark photon production in the early universe}
\label{sec:DarkPhotonProduction}
Now, we are ready for looking at the dark photon production in the early universe, where the dark photon can be a good candidate for the dark matter.
As the dark photon feebly couples to the SM particles due to the large $f_a$, it never reaches a thermal equilibrium in most parameter space; the dark photon is non-thermally produced.

\begin{figure}[t]
\subfigure[]{\label{fig:production1}
\includegraphics[width=0.23\textwidth,clip]{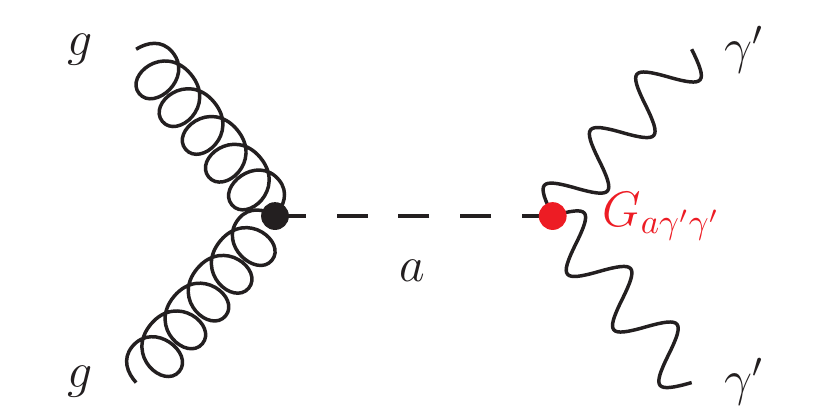}}
\subfigure[]{\label{fig:production2}
\includegraphics[width=0.23\textwidth,clip]{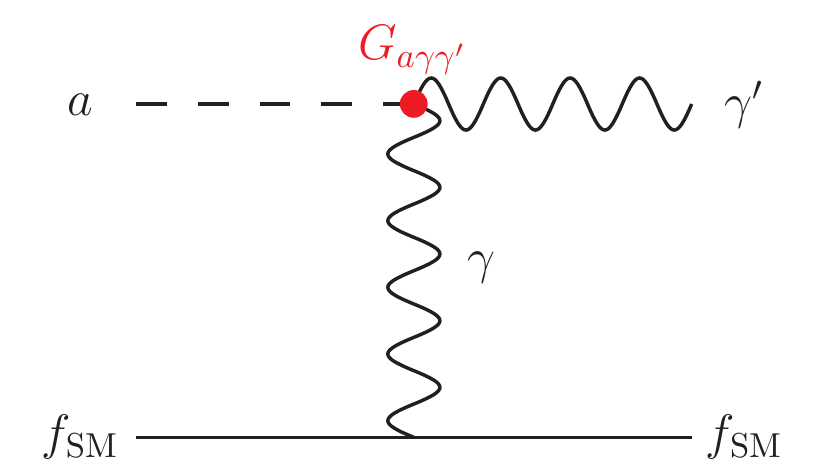}}
\caption{New production mechanisms of the dark photon dark matter using the dark axion portal in the early universe via (a) the axion mediation and (b) the dark Primakoff process, which is the dominant process for the case (i) $Q_\psi = 0$ and case (ii) $Q_\psi \ne 0$, respectively.}
\label{fig:productions}
\end{figure}

We will discuss case (i) and case (ii) in order as they employ quite a different production mechanism of the dark photon (see Fig.~\ref{fig:productions}).
It should be noted that there is another possible production process which is not related to the dark axion portal, namely, $gg\rightarrow \gamma'\gamma'$ process through the box diagram induced by the $\psi$ loop.
However, this process becomes negligible when the mass of $\psi$ is heavy, such as $f_a$, as long as we take $T<f_a$ during the $\gamma'$ production, by which the production rate of $\psi \bar\psi\to \gamma'\gamma'$ is naturally suppressed as well.
If the reheating temperature exceeds $f_a$, the PQ symmetry restoration may occur, which also leads to the vanishing dark axion portal.
Therefore, in the following discussion, we focus on the case where the reheating temperature does not exceed $f_a$ so that the PQ symmetry is never restored.

\subsection{Case (i)}
In case (i), the exotic colored fermions are electrically neutral, and the axion does not couple to the photon through the triangle diagram, which result in $G_{a\gamma\gamma} = G_{a\gamma\gamma'} = 0$.
Since $G_{a\gamma'\gamma'}$ is only the non-vanishing dark axion portal term, the dark photon can be produced via $gg \to a \to \gamma'\gamma'$ and becomes stable [Fig.~\ref{fig:production1}].

In Ref.~\cite{Kaneta:2016wvf}, the freeze-in mechanism \cite{Hall:2009bx} for the dark photon production via $gg\rightarrow a \rightarrow\gamma'\gamma'$ is analyzed.
The Boltzmann equation for the $\gamma'$ is given by
\bea
-sHT\frac{dY_{\gamma'}}{dT}=\gamma[n_{\gamma'}],
\label{eq:BoltzmannEQ}
\eea
where $Y_{\gamma'}=n_{\gamma'}/s$ is the comoving number density of the $\gamma'$, and $\gamma[n_{\gamma'}]$ denotes the collision term, $s=(2\pi^2/45)g_{*s}T^3$, $H^2=(\pi^2/90)g_{*\rho}T^4/M_{\rm Pl}^2$ with $M_{\rm Pl} \simeq 2.4\times10^{18}$ GeV, and $g_{*s}=g_{*\rho}\equiv g_*$ is taken as a constant value in our analysis.
The annihilation cross section of this process is $\sigma v \simeq 4G_{agg}G_{a\gamma'\gamma'}|A(\tau_\psi)|^2S$ with $A\equiv A(\tau_\psi)$ being the loop function given by
\bea
A\left(\tau_\psi\right) = \frac{1}{\tau_\psi}\left\{
\begin{tabular}{cc}
$\arcsin^2\sqrt{\tau_\psi}$ & $\tau_\psi \le 1$ \\
$-\frac{1}{4}\Big[\log\frac{1+\sqrt{1-\tau_\psi^{-1}}}{1-\sqrt{1-\tau_\psi^{-1}}} - i\pi\Big]^2$ & $\tau_\psi > 1$
\end{tabular}
 \right. , ~~
\eea
where $\tau_\psi\equiv S/(4m_\psi^2)$ with $m_\psi$ and $S$ being the mass of $\psi$ and the squared collision energy, respectively.
In the case that the $\psi$ is very heavy compared to the reheating temperature, we can take $A(\tau_\psi)\simeq 1$ in the thermally averaged cross section, and hereafter we restrict ourselves to this heavy $\psi$ case.
Then, the collision term in Eq. (\ref{eq:BoltzmannEQ}) is given by
\bea
\gamma_{gg\rightarrow \gamma'\gamma'} \simeq \frac{48}{\pi^4} G_{agg}^2G_{a\gamma'\gamma'}^2 T^8 \, ,
\eea
which leads to
\begin{eqnarray}
	\Omega_{\gamma'}h^2 &\simeq& 0.12\times g_D^4
	\Big(
		\frac{100}{g_*}
	\Big)^{3/2}
	\Big(
		\frac{m_{\gamma'}}{10~{\rm keV}}
	\Big)\nonumber\\
	&&
	\times
	\Big(
		\frac{5T_{\rm RH}}{f_a}
	\Big)^3
	\Big(
		\frac{10^{10}~{\rm GeV}}{f_a}
	\Big) ,
\end{eqnarray}
where $m_{\gamma'}$ and $T_{\rm RH}$ are the mass of the dark photon and the reheating temperature, respectively, and we define $g_D \equiv e' D_\psi/0.3$.

\begin{center}
\begin{figure}[t]
\includegraphics[width=0.43\textwidth]{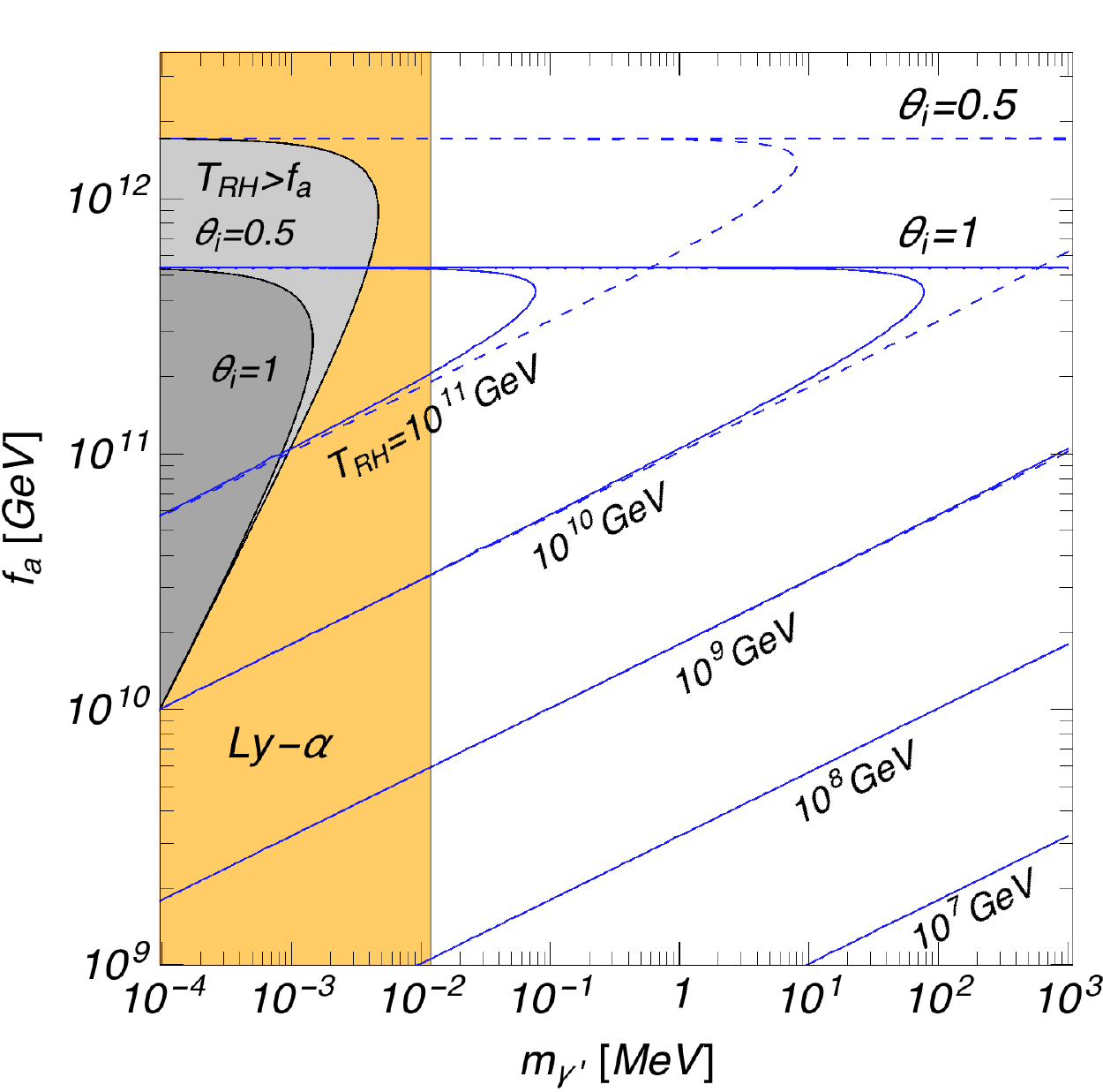}
\caption{
The blue curves show $\Omega_{\rm DM}h^2 = (\Omega_{\gamma'} + \Omega_a) h^2 = 0.12$ for the given $T_{\rm RH}$ with the initial misalignment angle $\theta_i = 0.5$ (blue dashed curves) and $\theta_i=1$ (blue solid curves) in the case (i) for a choice of $g_D = 1$.
The gray regions are disfavored, since the reheating temperature to obtain the correct dark matter density exceeds $f_a$, restoring the PQ symmetry. The orange region shows the Lyman-$\alpha$ constraint only for the case of $\Omega_{\rm DM}h^2=\Omega_{\gamma'}h^2$.
}
\label{fig:AMWithAxionCDM}
\end{figure}
\end{center}

The observed dark matter number density is accounted for by the axion and dark photon together, $\Omega_{\rm DM}h^2 = (\Omega_{\gamma'} + \Omega_a) h^2 = 0.12$, and it is shown in Fig. \ref{fig:AMWithAxionCDM} for $g_D=1$.
In the figure, the blue solid and dashed curves become horizontal in the large $f_a$ region, since the whole amount of the dark matter density can be explained by the axion alone, while in the smaller $f_a$ region the dark photon can compensate the shortage of the axion dark matter.
In the case of $\Omega_{\rm DM}h^2=\Omega_{\gamma'}h^2$, the dark photon lighter than $O(1 - 10)$ keV may affect the small scale structure, and thus the Lyman-$\alpha$ forest gives a lower limit on $m_{\gamma'}$, depicted by the orange region in Fig.~\ref{fig:AMWithAxionCDM}, which we take $m_{\gamma'}\lesssim 12$ keV \cite{Baur:2015jsy}.
\footnote{Although a careful analysis of the power spectrum is needed for more accurate constraint, we have taken this crude value by maintaining the entropy density for the early-decoupled dark matter case.}
It should be noted that there is an upper bound on $T_{\rm RH}$, since if $T_{\rm RH}$ was sufficiently high, $\gamma'$ could have been thermalized and produced too much to explain the observed value.
By demanding $H(T_{\rm RH})<\gamma_{gg\to\gamma'\gamma'}(T_{\rm RH})/n_{\gamma'}^{\rm eq}(T_{RH})$, with $n_{\gamma'}^{\rm eq} \simeq [3 \zeta(3) / \pi^2] T^3$, we obtain
\bea
T_{\rm RH}\lesssim 10^{10}~{\rm GeV} \times g_D^{4/3}\Big(\frac{g_*}{100}\Big)^{1/6}\Big(\frac{f_a}{10^{10}~{\rm GeV}}\Big)^{4/3} ~
\label{eq:nonthermalCase1}
\eea
which can be satisfied by taking a larger $m_{\gamma'}$ for the dark matter abundance to be the observed value (see Fig. \ref{fig:AMWithAxionCDM}).

\subsection{Case (ii)}

In case (ii), since the $\psi$ is electrically charged, the $G_{a\gamma\gamma'}$ does not vanish, which leads to the decay of the $\gamma'$ into the axion and photon.
The partial decay width of $\gamma'\rightarrow a\gamma$, which is the main decay channel, is given by
\bea
\Gamma (\gamma' \to \gamma a) = \frac{G_{a \gamma \gamma'}^2}{96 \pi} m_{\gamma'}^3 \Big[ 1 - \frac{m_a^2}{m_{\gamma'}^2} \Big]^3 . \label{eq:gammaPrimeDecay}
\eea
For small $m_{\gamma'}$ and/or large $f_a$, the dark photon becomes sufficiently long-lived particles so that it can be a dark matter.

At higher temperature ($T\gg v_{\rm EW}$), the axion is also thermalized through $gg\leftrightarrow ga$ and other hadronic processes \cite{Masso:2002np,Graf:2010tv,Salvio:2013iaa}.
Therefore, the dark photon production through the dark Primakoff process, $fa\to f\gamma'$ with $\gamma'$ being non-thermal, becomes efficient, which is similar to the thermal axion production in the electron-photon scattering ($\gamma e \rightarrow a e$) \cite{Bolz:2000fu}.

The $G_{a\gamma\gamma'}$ also contributes to another $\gamma'$ abundance produced by the annihilation of the SM particles ($f\bar{f}\rightarrow \gamma\rightarrow a\gamma'$). Since the $\gamma$ involved in this s-channel process has the thermal mass, the plasmon decay takes place if the temperature is high enough. Compared to the dark Primakoff process, however, the plasmon decay contribution is negligible \cite{Fukugita:1982ep}.

It would be worthwhile to note that if there are direct couplings between $\gamma'$ and the SM fermion through, for instance, the kinetic mixing, the t-channel annihilation process, $f \bar f \to \gamma' V$ with $V$ being the SM gauge bosons, may also give a significant contribution at high temperature, \cite{Arias:2012az} which we have omitted by turning off the kinetic mixing.

The collision term of the dark Primakoff process is given by
\bea
\gamma_{fa\to f\gamma'} \simeq g_F(T)\frac{T^6}{\pi^4} \frac{e^2G_{a\gamma\gamma'}^2}{8\pi}\left(\log \frac{T^2}{m_\gamma^2}+\alpha_{\gamma'}\right)
\label{eq:CollisionTermsDarkPrimakoff}
\eea
where $g_F\equiv \sum_f g_{f}Q_f^2$ counts the number of the relativistic degrees of freedom $g_f$ of electrically charged fermions at the temperature $T$, and  $\alpha_{\gamma'} = 3/4 - 2\gamma_E + \log 4$ with $\gamma_E$ being Euler's constant given by $\gamma_E\simeq 0.5772$. For more detail of Eq.~\eqref{eq:CollisionTermsDarkPrimakoff}, see Appendix~\ref{sec:appB}.
Here, we have introduced the photon thermal mass $m_\gamma \sim eT$ to regulate the infrared divergence.\footnote{While this cutoff method is often used (e.g., Ref.~\cite{Pilaftsis:2003gt,Kawasaki:2004qu}), a more accurate treatment exists \cite{Bolz:2000fu}. The cutoff method, however, provides the resultant reaction rate at the same order of magnitude as that obtained by the more rigorous calculation, and this is good enough for our purpose in this paper.}
By integrating $dY_{\gamma'}/dT$ over $T$ from $T=T_{\rm RH}$ to $T\simeq 0$ in Eq.~\eqref{eq:BoltzmannEQ}, we obtain
\bea
	Y_{\gamma'}^0
	& \simeq &
	\frac{135\sqrt{10}}{2\pi^7g_*^{3/2}}\frac{4e^2G_{a\gamma\gamma'}^2}{\pi}M_{\rm Pl}T_{\rm RH}\nonumber\\
	&& \times \frac{g_F\left(T_{\rm RH}\right)}{32} \left(\log\frac{T^2}{m_\gamma^2}+\alpha_{\gamma'}-2\right),
\eea
and thus the dark photon abundance is given by
\bea
\Omega_{\gamma'}h^2 
& \simeq  0.12 \times \bar g_D^2\left(\frac{Q_\psi}{1/3}\right)^2\left(\frac{100}{g_*}\right)^{3/2}\left(\frac{g_F\left(T_{\rm RH}\right)}{32}\right) 
\nonumber\\
& \times\left(\frac{m_{\gamma'}}{{\rm MeV}}\right)\left(\frac{10^2}{\vpq/T_{RH}}\right)\left(\frac{10^{10}}{\vpq/{\rm GeV}}\right)
\eea
with $\bar g_D \equiv e' D_\psi/0.01$ .

In addition to the dark Primakoff process, there is another contribution from $gg\to\gamma'\gamma'$ as discussed in case (i).
However, because of the difference in $\gamma_{fa\to f\gamma'} \propto 1/f_a^2$ and $\gamma_{gg\to\gamma'\gamma'} \propto 1/f_a^4$, the dark photon production from the dark Primakoff process is dominant contribution in the parameter space of our interest.

\begin{widetext}
\begin{center}
\begin{figure}[t]
\subfigure[]{
\includegraphics[width=0.32\textwidth]{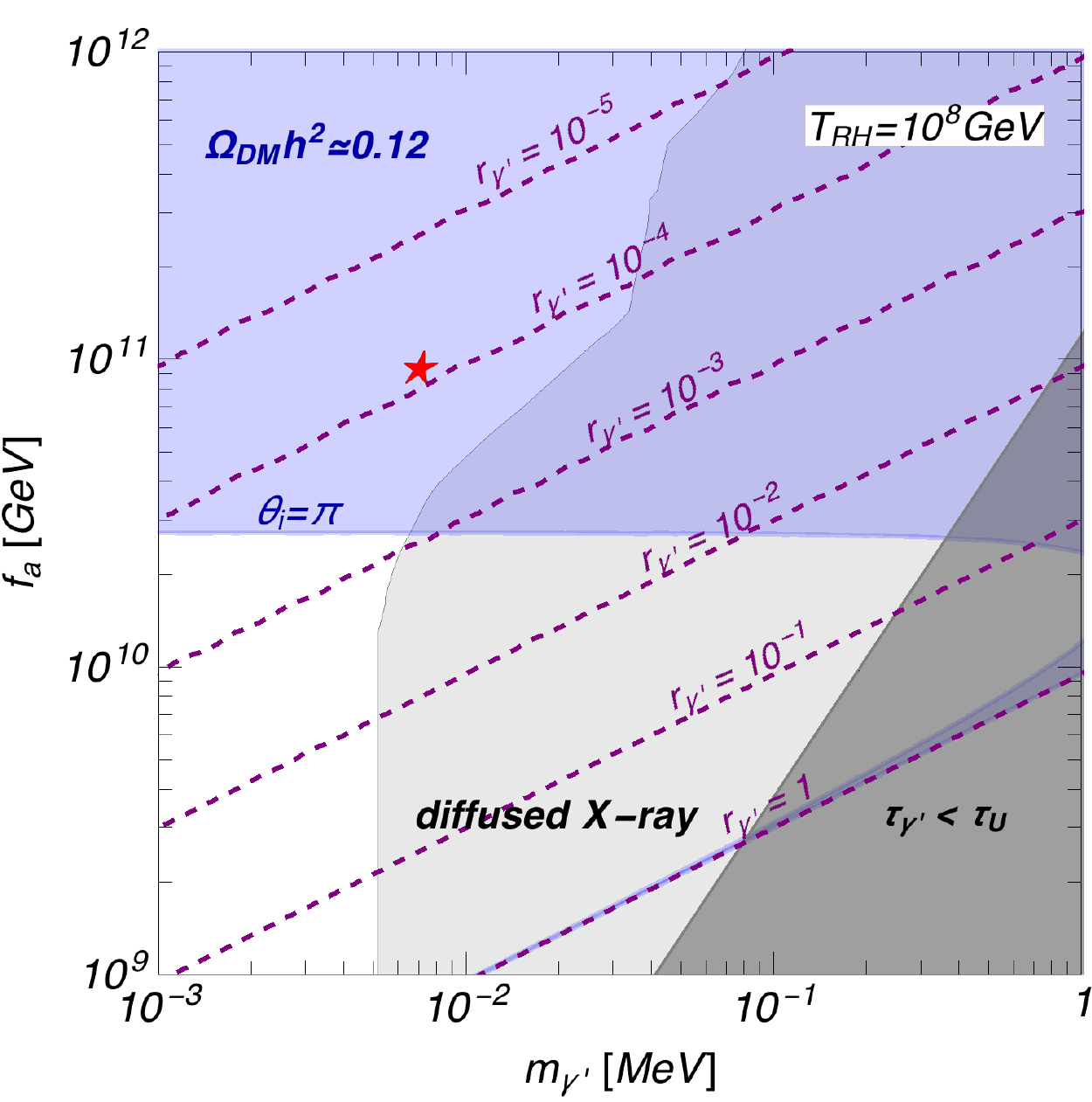}}
\subfigure[]{
\includegraphics[width=0.32\textwidth]{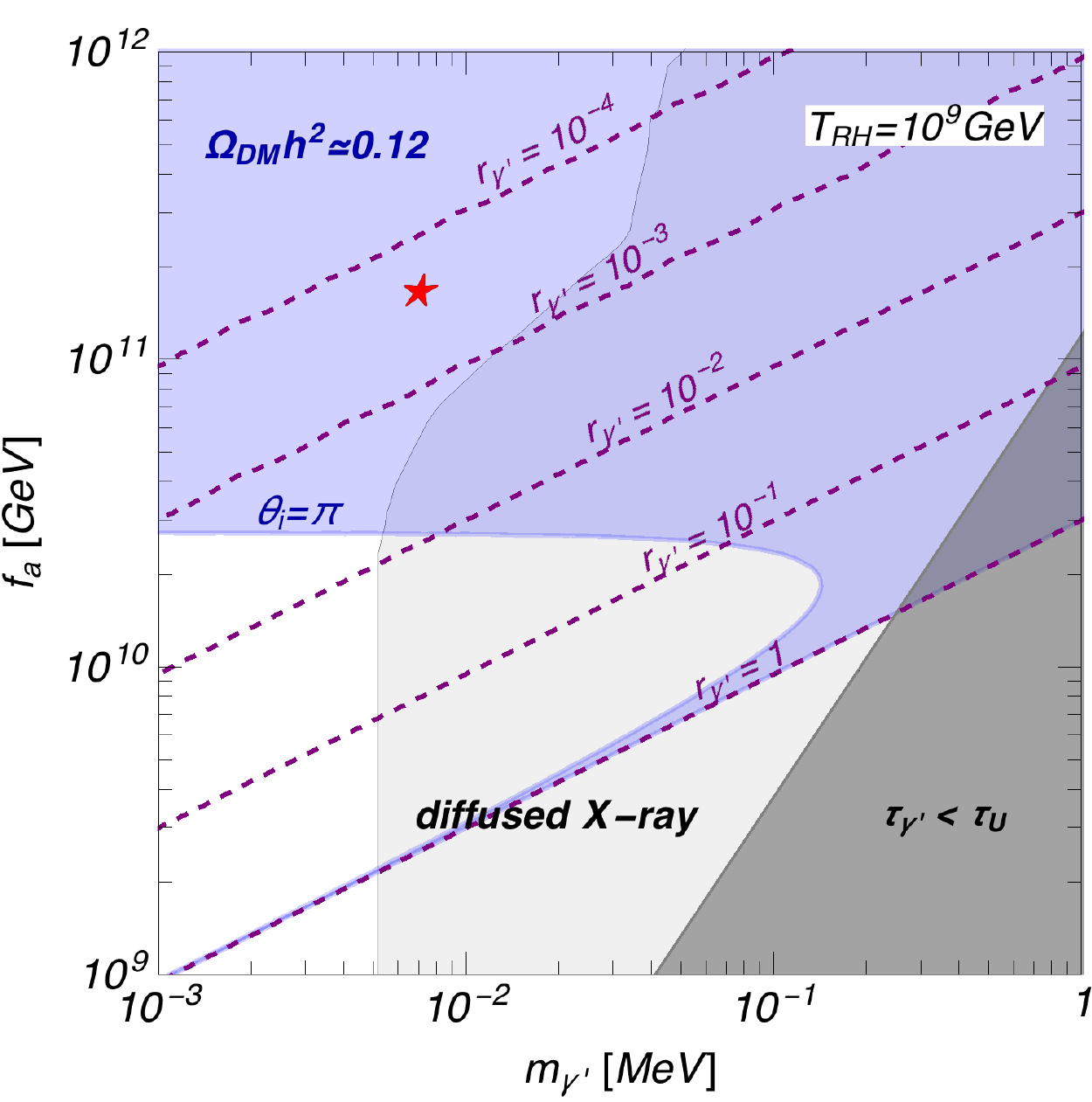}}
\subfigure[]{
\includegraphics[width=0.32\textwidth]{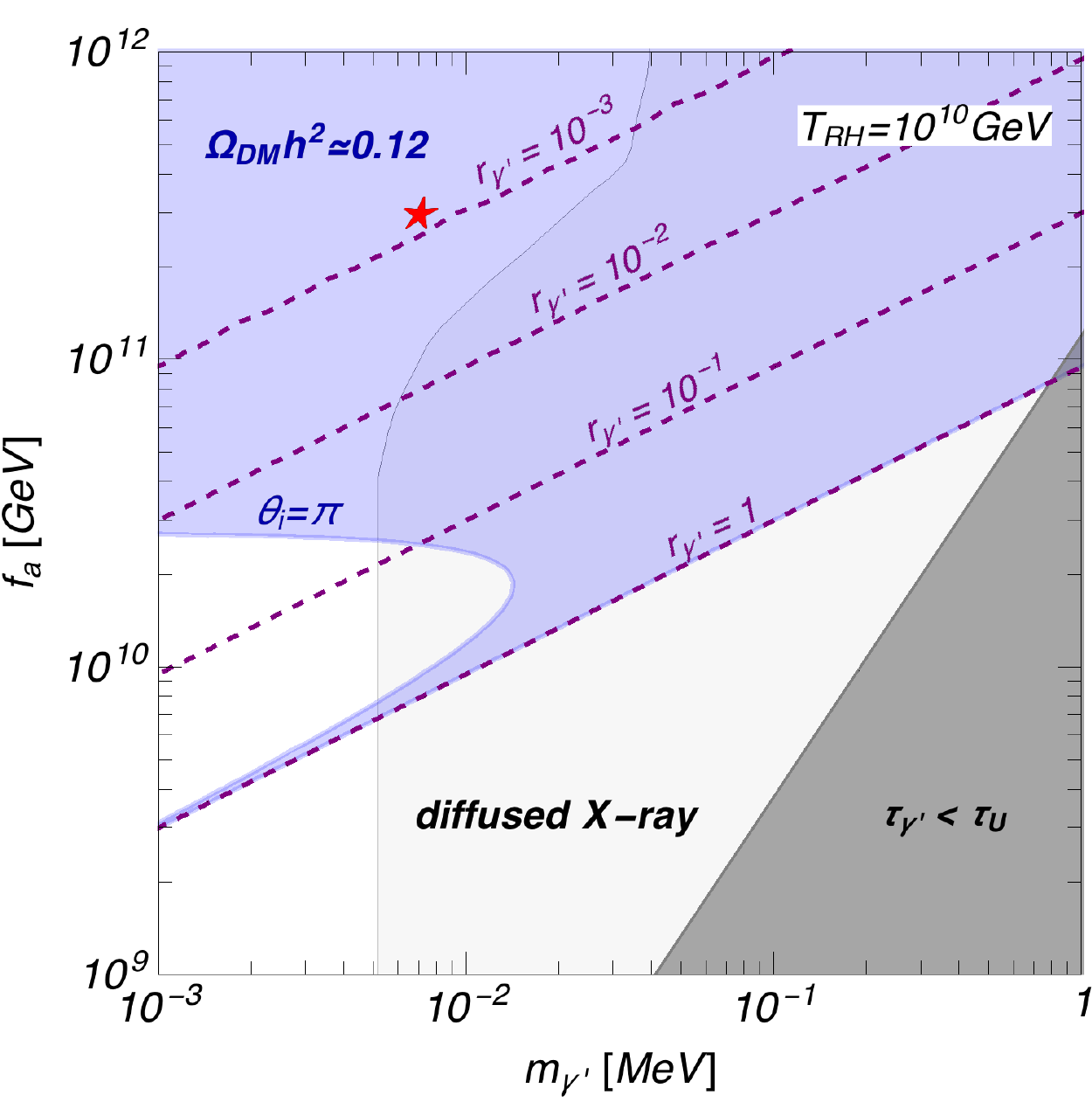}}
\caption{
The blue regions show $\Omega_{\rm DM}h^2 = (\Omega_{\gamma'} + \Omega_a) h^2  = 0.12$  with the initial misalignment angle varying from $\theta_i$ from 0 to $\pi$ in the case (ii) for a choice of $\bar g_D = 1$.  It depends on the reheating temperature, and we illustrate for $T_{\rm RH} =$ (a) $10^8~{\rm GeV}$, (b) $10^9~{\rm GeV}$, and (c) $10^{10}~{\rm GeV}$. The dark and light gray regions depict the bounds from $\tau_{\gamma'} < \tau_U$ and the observations of the diffused $X$-ray, respectively. The red stars indicate the points that can explain the 3.5 keV $X$-ray excess via the $\gamma' \to \gamma a$ decay.
}
\label{fig:PrimakoffDarkPhotonWithCDMAxion}
\end{figure}
\end{center}
\end{widetext}

Figure \ref{fig:PrimakoffDarkPhotonWithCDMAxion} shows the regions of $\Omega_{\rm DM}h^2=0.12$ with $\theta_i$ varying form 0 to $\pi$, for a choice of $\bar g_D=1$.
Each panel of the figure shows a different $T_{\rm RH}$ value case.
The dark gray regions in the figure represent the case that the lifetime of the dark photon, $\tau_{\gamma'}$, becomes shorter than the age of the universe, $\tau_U\simeq 13.7\times10^9$ yrs.
As the dark photon decays into a photon in the case (ii), the non-observation of the dark matter signal in the diffused $X$-ray gives a stronger constraint on the lifetime of the dark photon.
A generic constraint on the dark matter lifetime is discussed in Ref. \cite{Essig:2013goa} from which we have estimated the bound in our case.
By taking into account the dependence on
\bea
r_{\gamma'} \equiv \frac{\Omega_{\gamma'}h^2}{\Omega_{\rm DM}h^2}
\eea
in the diffused $X$-ray flux, we show its bound as light gray regions in the figure.
The point noted in the red star in the figure indicates that the dark photon can explain the 3.5 keV $X$-ray line excess, which we will discuss in the next section.

It should be noted that there is an upper bound on $T_{\rm RH}$ for the dark photon to remain non-thermal in a similar way to Eq.~\eqref{eq:nonthermalCase1}.
\bea
T_{\rm RH} \lesssim (3 \times 10^{11}~{\rm GeV}) \, \bar g_D^{-2}\left(\frac{g_*}{100}\right)^{1/2}\left(\frac{f_a/{\rm GeV}}{10^{10}}\right)^{2} . ~~~
\label{eq:UpperLimitReheatingThermal}
\eea
In addition, for the dark Primakoff process to be effective, the axion should be still in thermal bath during the dark photon production.
The thermalization of the axion is maintained by the reaction $gg\leftrightarrow ga$ whose decoupling temperature $T_D$ is given by \cite{Salvio:2013iaa}
\bea
T_D \simeq (10^5 ~ {\rm GeV}) \left(\frac{\vpq}{10^{10} ~ {\rm GeV}}\right)^2,
\label{eq:LowerLimitReheating}
\eea
and thus $T_{\rm RH}>T_D$ should be satisfied.
Therefore, in the case of $\Omega_{\gamma'}\gg\Omega_a$, i.e. $\Omega_{\gamma'}h^2 \simeq 0.12$, the conditions (\ref{eq:UpperLimitReheatingThermal}) and $T_{\rm RH}>T_D$ respectively give the lower and upper bounds on $m_{\gamma'}$, and we obtain $1~{\rm keV}\lesssim m_{\gamma'}\lesssim 1~{\rm GeV}\times \bar g_D^{-2}$.
The regions realizing $\Omega_{\gamma'}\gg\Omega_a$ are, however, disfavored by the diffused $X$-ray bound and the requirement of $T_{\rm RH}\ll f_a$, as shown in Fig.~\ref{fig:PrimakoffDarkPhotonWithCDMAxion}.

\section{\boldmath Dark photon explanation of the 3.5 keV $X$-ray line excess}
\label{sec:3.5keV}
In various $X$-ray observations of galaxy clusters, it has been observed that there is an anomalous excess at $3.5 ~{\rm keV}$ in the $X$-ray spectra from galaxies. \cite{Bulbul:2014sua,Boyarsky:2014jta,Riemer-Sorensen:2014yda,Jeltema:2014qfa,Boyarsky:2014ska,Iakubovskyi:2015dna}, and thus it is worth discussing whether it can be explained by the dark photon dark matter in our scenario.

It is known that the dark matter mass ($m_{\rm DM}$) and its lifetime ($\tau_{\rm DM}$) (with a decay to a photon final state) are required to be~\cite{Boyarsky:2014jta}
\bea
m_{\rm DM} & \simeq & 7 ~{\rm keV} , \label{eq:massCondition} \\
\tau_{\rm DM} & \simeq & 10^{28} ~ {\rm sec} \simeq 3 \times 10^{20} ~ {\rm yrs} \, ,
\label{eq:DMLifeTimeXray}
\eea
so that the particle dark matter can explain the 3.5 keV $X$-ray line.\footnote{Although the tension between such a light dark matter and the constraint from the small scale structure is currently under debate, the tension is ameliorated if the 7 keV dark matter is sub-dominant component of the whole dark matter abundance \cite{Baur:2015jsy,Boyarsky:2008mt,Harada:2014lma,Kamada:2016vsc}, which is the case we discuss.}
The case (i) does not have the dark photon decay mode to the photon while the case (ii) does.
In the following discussion, we consider only the $\gamma' \to \gamma a$ in the case (ii) and take $G_{a\gamma\gamma'}$ as a free parameter instead of specifying $e'$ and $Q_\psi$.

In case (ii), the lifetime of the dark photon is given by
\begin{eqnarray}
	\tau_{\gamma'} &\simeq& (1.5\times10^{25}~{\rm sec})
	\left(
		\frac{10^{-16}~{\rm GeV}^{-1}}{G_{a\gamma\gamma'}}
	\right)^2
	\left(
		\frac{7~{\rm keV}}{m_{\gamma'}}
	\right)^3 \label{eq:gammaPrimeLifetime}
\end{eqnarray}
from Eq.~\eqref{eq:gammaPrimeDecay}, which is shorter than the condition (\ref{eq:DMLifeTimeXray}).
On the other hand, when the dark photon is responsible of only a fraction of the total dark matter abundance, the condition changes to \cite{Demidov:2014hka}
\begin{eqnarray}
	\tau_{\rm DM} &\simeq& r_{\gamma'} \times 10^{28}~{\rm sec} , \label{eq:modifiedCondition}
\end{eqnarray}
where $r_{\gamma'}$ can be written in terms of $G_{a\gamma\gamma'}$ as
\begin{eqnarray}
	r_{\gamma'} &\simeq& 10^{-3}
	\left(
		\frac{100}{g_*}
	\right)^{3/2}
	\left(
		\frac{T_{\rm RH}}{10^{11}~{\rm GeV}}
	\right)\nonumber\\
	&&\times
	\left(
		\frac{G_{a\gamma\gamma'}}{10^{-16}~{\rm GeV}^{-1}}
	\right)^2
	\left(
		\frac{m_{\gamma'}}{7~{\rm keV}}
	\right),
\end{eqnarray}
and thus the dark photon can produce the observed 3.5~keV $X$-ray line if $\tau_{\gamma'}$ of Eq.~\eqref{eq:gammaPrimeLifetime} satisfies the condition \eqref{eq:modifiedCondition}.

\begin{center}
\begin{figure}[t]
\includegraphics[width=0.43\textwidth]{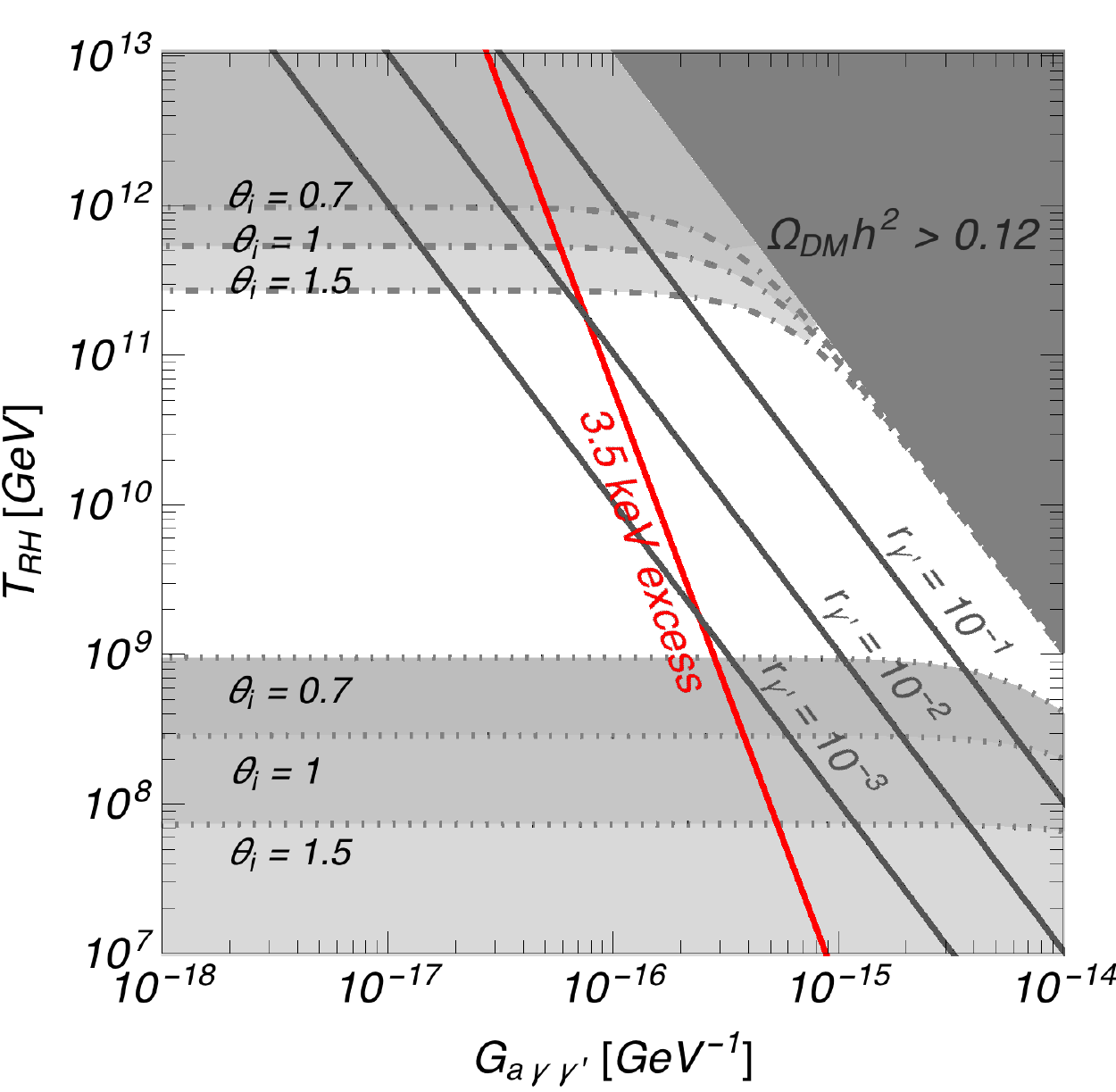}
\caption{Parameter space of the 2-component dark matter scenario where the axion CDM is dominant and the $7 ~\kev$ dark photon produced by the dark Primakoff mechanism is sub-dominant (with a fraction of $r_{\gamma'}$).
The red line can explain the 3.5 ${\rm keV}$ $X$-ray line signal.
The shaded regions are constrained by the PQ symmetry restoration condition (dot-dashed bounds), the overproduction of the dark photon dark matter (dark gray corner), and the axion thermalization condition (dotted bounds).
$\theta_i$ is the initial misalignment angle on which the axion relic density depends.
}
\label{fig:XrayLineSignal}
\end{figure}
\end{center}

In Fig.~\ref{fig:XrayLineSignal}, we fix $m_{\gamma'}=7~{\rm keV}$ as required by the condition \eqref{eq:massCondition}.
The red line shows the parameter region that can explain the 3.5 keV $X$-ray excess by satisfying the condition \eqref{eq:modifiedCondition}, while the gray solid lines correspond to several values of $r_{\gamma'}$.
TThe 3.5 keV solution provides a relation between the $G_{a\gamma\gamma'}$ and $T_\text{RH}$ as
\bea
G_{a\gamma\gamma'} \simeq (10^{-16} ~\text{GeV}^{-1}) \left( \frac{10^{11} ~\text{GeV}}{T_\text{RH}} \right)^{1/4}
\eea
or
\bea
\frac{f_a}{PQ_\Phi} \simeq (10^{12} ~\text{GeV}) \left| \frac{e' D_\psi}{0.01} \frac{Q_\psi}{1/3} \right| \left(\frac{T_\text{RH}}{10^{11} ~\text{GeV}}\right)^{1/4} ,
\eea
which is illustrated as the $T_\text{RH}$ increases in Fig.~\ref{fig:PrimakoffDarkPhotonWithCDMAxion}.

The condition $\Omega_{\rm DM} h^2 = 0.12$ holds in the whole region in the figure, except for the dark gray corner where the dark photon is overproduced ($\Omega_{\rm DM}h^2\sim \Omega_{\gamma'}h^2 > 0.12$). 

The lightly shaded regions of the parameter space, while satisfying the relic density condition, are disfavored by other considerations.
The upper light gray regions (dot-dashed boundary) for $G_{a\gamma\gamma'} \lsim 10^{-15} ~{\rm GeV}^{-1}$, do not satisfy the upper limit on $T_{\rm RH}$ given by the PQ symmetry restoration condition ($T_{\rm RH} < f_a$).
For $G_{a\gamma\gamma'} \gsim 10^{-15} ~{\rm GeV}^{-1}$, the dark photon non-thermalization condition to avoid DM overproduction of Eq.~\eqref{eq:UpperLimitReheatingThermal} engages.
The lower light gray regions (dotted boundary) are disfavored, since the axion thermalization condition ($T_{\rm RH} >T_D$) is not satisfied.
The $\theta_i$ dependence of all the light gray regions is because $f_a$ is determined by $\theta_i$ when we demand $\Omega_{\rm DM}h^2\sim \Omega_ah^2 = 0.12$ in the region of $r_{\gamma
'}\ll 1$.

\section{Discussions on some issues}
\label{sec:discussions}
In this section, we have brief discussions on the several issues of the dark axion portal although they are not our main focus in our paper.

First, we discuss the direct detection of the dark photon dark matter.
The axion relic dark matter is searched for using the $G_{a\gamma\gamma}$ coupling \cite{Asztalos:2009yp,CAPP}.
Assuming the dark photon dark matter makes up a large fraction of the total dark matter relic density, we briefly comment on its detection possibility.
If the kinetic mixing $\varepsilon$ is large enough, the axion dark matter experiments \cite{Asztalos:2009yp,CAPP} may be used to search for a dark photon dark matter of a similar mass ($m_{\gamma'} \approx 10^{-6} - 10^{-4}$ eV)\cite{Wagner:2010mi}.
When we ignore the kinetic mixing effect and consider only the $G_{a\gamma\gamma'}$ coupling, the dark photon may be searched for using a scatterting with an electron inside the detector.
The scattering would be mediated by the photon and the dark photon would convert into an axion which would escape the detector.
A low-$Q^2$ electron recoil would be a signal of such a dark photon dark matter using the dark axion portal.
It would require a careful analysis to see if an existing dark matter detection experiment can detect it or a new detector design is necessary.

Now, we remind ourselves the stability issue of the exotic quark ($\psi$, $\psi^c$).
The exotic quark in the original KSVZ model does not decay at all, but it may decay in the dark KSVZ model depending on the charge assignment because of an additional particle $\Phi_D$.
For instance, $\Phi_D^\dagger \psi D^c$ term is allowed for $PQ_\psi = 0$, $Q_\psi = -1/3$, $D_\psi = D_\Phi$ in the case (ii), which would allow the exotic quark decay into the $S_D$ (CP-even component of the $\Phi_D$) and a down-type quark \cite{Kaneta:2016wvf}.
This term might cause a flavor-changing neutral current such as $b \to s + \gamma'$, but it will be highly suppressed by the large mass of the exotic quark.
In the case (i), we do not have such a decay mode, yet we can adopt the same attitude to this issue as the original KSVZ model.
As the exotic quark mass scale ($f_a \gsim 10^9 ~\gev$) is larger than the reheating temperature, the exotic quarks would have been produced too little in the early universe to cause any conflict with the experimental data.

Finally, let us comment on a possible extension of the model and its phenomenology.
Though we have limited ourselves to the conventional QCD axion case in this paper where $m_a \sim (10^{-5} - 10^{-2}) ~\ev$ and $G_{a\gamma\gamma} \sim (10^{-11} - 10^{-14}) ~\gev^{-1}$, our discussion can be extended to rather a wide class of models with an axion-like particle (ALP) whose mass and coupling can be significantly larger than the QCD axion. 
In such models, various types of experiments can be used to test the dark axion portal, perhaps in a similar way that the colliders give constraints on the axion-photon-$Z$ boson coupling by measuring $Z\to 3 \gamma$ following $Z\to \gamma + {\rm ALP}$ for a MeV-GeV scale ALP~\cite{Jaeckel:2015jla,Alves:2016koo}.
For instance, a mono-photon signal from $Z\to\gamma + 2\gamma'$ followed by $Z\to \gamma + {\rm ALP}$ and $gg\to {\rm ALP} \to \gamma \gamma'$ would be possible.
Using the similar diagrams, a mono-$Z$ signal would be also possible.
As other interesting channels, $Z\to 2\gamma + \gamma'$ induced by the exotic fermion loop and/or followed by ${\rm ALP}\to \gamma+\gamma'$ might be also worthwhile studying in the future.

\section{Summary}
\label{sec:summary}
In this paper we have discussed a new mechanism to produce the dark photon in the early universe with the help of the axion using the dark axion portal.
In particular, the dark photon can be a dark matter candidate if its lifetime is long enough.

Our discussion is categorized in two cases based on the dark KSVZ model: (i) the vector-like fermion is electrically neutral, (ii) it is electrically charged.
In both cases the dark photon can have longer lifetime than the age of the universe.
On the other hand, the dark photon production process is quite different in each case, while the dark photon is always non-thermal relic due to a feeble coupling to the SM particles.
In case (i), $gg\to a\to \gamma'\gamma'$ via $G_{a \gamma' \gamma'}$ is the only possible way to produce the dark photon unless we count on the kinetic mixing.
Interestingly, even in the small $f_a$ regions where the axion abundance is sufficiently small, the dark photon can compensate the dark matter abundance to achieve the observed value.
The dark Primakoff process, $f a \to f \gamma'$ via $G_{a \gamma \gamma'}$, is open in case (ii) to produce the dark photon.
Since the dark photon decays into a photon in this case, we have discussed the diffused $X$-ray constraint.
It can also explain the 3.5 keV $X$-ray line excess as we discussed.

\begin{acknowledgments}
This work was supported by IBS (Project Code IBS-R018-D1).
HL appreciates hospitality during his visit to KIAS.
We thank many people including KJ Bae, EJ Chun and P Ko for intriguing discussions, and A Kamada for helpful comments.
\end{acknowledgments}

\appendix

\section{Collision term of the dark Primakoff process}
\label{sec:appB}
Here we give more detailed expression related to the collision term of the dark Primakoff process in Eq.~\eqref{eq:CollisionTermsDarkPrimakoff}.
Let us consider the process $f a \to f \gamma'$ by exchanging single photon, where $f$ denotes a SM fermion having electric charge $Q_f$.
The squared amplitude after summing over the spin is then given by
\bea
\left|\mathcal{M_{\mathit{f}}}\right|^2 \simeq \frac{e^2Q_f^2G_{a\gamma\gamma'}^2}{2} \frac{\left(-T^3 - 2S^2T - 2ST^2\right)}{\left(T-m_\gamma^2\right)^2},
\eea
where $S$ and $T$ are the Mandelstam variables, and $m_{\gamma'} \sim e T$ is the plasmon mass.
We have taken all the external particles to be massless.
The collision term in the Boltzmann equation can be written as
\begin{eqnarray}
	\gamma_{fa \to f\gamma'}
	&=&
	\sum_fg_f\frac{T}{32\pi^4}\int dS \, (\sigma_f v) \, S^{3/2} K_1\Big(\frac{\sqrt{S}}{T}\Big) , ~~
\end{eqnarray}
where $g_f$ is the number of degrees of freedom of particle $f$, the cross section is given by
\begin{eqnarray}
	\sigma_f v &\simeq&
	\frac{e^2Q_f^2G_{a\gamma\gamma'}^2}{32\pi}
	\left(
		4\log\frac{S}{m_{\gamma}^2} - 7
	\right) ,
\end{eqnarray}
and thus we end up with
\bea
\gamma_{fa \to f\gamma'} \simeq g_F(T) \frac{T^6}{\pi^4} \frac{e^2G_{a\gamma\gamma'}^2}{8\pi}\left(\log \frac{T^2}{m_\gamma^2}+\alpha_{\gamma'}\right)
\eea
with $g_F(T) \equiv \sum_f g_f Q_f^2$ at the relevant temperature and $\alpha_{\gamma'} = 3/4 - 2\gamma_E + \log 4$ .




\begin{thebibliography}{99}
\bibitem{Gershtein:2013iqa} 
  Y.~Gershtein {\it et al.},
  arXiv:1311.0299 [hep-ex].

\bibitem{Essig:2013lka}
  R.~Essig {\it et al.},
  arXiv:1311.0029 [hep-ph].

\bibitem{Kusenko:2013saa} 
  A.~Kusenko and L.~J.~Rosenberg,
  arXiv:1310.8642 [hep-ph].
  
  \bibitem{Cushman:2013zza} 
  P.~Cushman {\it et al.},
  arXiv:1310.8327 [hep-ex].
  
\bibitem{Tan:2016zwf} 
  A.~Tan {\it et al.} [PandaX-II Collaboration],
  Phys.\ Rev.\ Lett.\  {\bf 117}, no. 12, 121303 (2016)
  doi:10.1103/PhysRevLett.117.121303
  [arXiv:1607.07400 [hep-ex]].
  
  \bibitem{Akerib:2016vxi} 
  D.~S.~Akerib {\it et al.} [LUX Collaboration],
  Phys.\ Rev.\ Lett.\  {\bf 118}, no. 2, 021303 (2017)
  doi:10.1103/PhysRevLett.118.021303
  [arXiv:1608.07648 [astro-ph.CO]].
  
  \bibitem{Alexander:2016aln} 
  J.~Alexander {\it et al.},
  arXiv:1608.08632 [hep-ph].

\bibitem{Kaneta:2016wvf} 
  K.~Kaneta, H.~S.~Lee and S.~Yun,
  Phys.\ Rev.\ Lett.\  {\bf 118}, no. 10, 101802 (2017)
  doi:10.1103/PhysRevLett.118.101802
  [arXiv:1611.01466 [hep-ph]].

\bibitem{Berezhiani:2000gh} 
  Z.~Berezhiani, L.~Gianfagna and M.~Giannotti,
  Phys.\ Lett.\ B {\bf 500}, 286 (2001)
  doi:10.1016/S0370-2693(00)01392-7
  [hep-ph/0009290].

\bibitem{Ejlli:2016asd} 
  D.~Ejlli,
  arXiv:1609.06623 [hep-ph].
      
\bibitem{Choi:2016kke} 
  K.~Choi, H.~Kim and T.~Sekiguchi,
  Phys.\ Rev.\ D {\bf 95}, no. 7, 075008 (2017)
  doi:10.1103/PhysRevD.95.075008
  [arXiv:1611.08569 [hep-ph]].
      
\bibitem{Bulbul:2014sua} 
  E.~Bulbul, M.~Markevitch, A.~Foster, R.~K.~Smith, M.~Loewenstein and S.~W.~Randall,
  Astrophys.\ J.\  {\bf 789}, 13 (2014)
  doi:10.1088/0004-637X/789/1/13
  [arXiv:1402.2301 [astro-ph.CO]].
  
\bibitem{Boyarsky:2014jta} 
  A.~Boyarsky, O.~Ruchayskiy, D.~Iakubovskyi and J.~Franse,
  Phys.\ Rev.\ Lett.\  {\bf 113}, 251301 (2014)
  doi:10.1103/PhysRevLett.113.251301
  [arXiv:1402.4119 [astro-ph.CO]].
  
\bibitem{Riemer-Sorensen:2014yda} 
  S.~Riemer-Sørensen,
  Astron.\ Astrophys.\  {\bf 590}, A71 (2016)
  doi:10.1051/0004-6361/201527278
  [arXiv:1405.7943 [astro-ph.CO]].
  
\bibitem{Jeltema:2014qfa} 
  T.~E.~Jeltema and S.~Profumo,
  Mon.\ Not.\ Roy.\ Astron.\ Soc.\  {\bf 450}, no. 2, 2143 (2015)
  doi:10.1093/mnras/stv768
  [arXiv:1408.1699 [astro-ph.HE]].
  
\bibitem{Boyarsky:2014ska} 
  A.~Boyarsky, J.~Franse, D.~Iakubovskyi and O.~Ruchayskiy,
  Phys.\ Rev.\ Lett.\  {\bf 115}, 161301 (2015)
  doi:10.1103/PhysRevLett.115.161301
  [arXiv:1408.2503 [astro-ph.CO]].
  
\bibitem{Iakubovskyi:2015dna} 
  D.~Iakubovskyi, E.~Bulbul, A.~R.~Foster, D.~Savchenko and V.~Sadova,
  arXiv:1508.05186 [astro-ph.HE].
  
\bibitem{Baker:2006ts} 
  C.~A.~Baker {\it et al.},
  Phys.\ Rev.\ Lett.\  {\bf 97}, 131801 (2006)
  doi:10.1103/PhysRevLett.97.131801
  [hep-ex/0602020].
  
\bibitem{Peccei:1977hh} 
  R.~D.~Peccei and H.~R.~Quinn,
  Phys.\ Rev.\ Lett.\  {\bf 38}, 1440 (1977).
  doi:10.1103/PhysRevLett.38.1440

\bibitem{Peccei:1977ur} 
  R.~D.~Peccei and H.~R.~Quinn,
  Phys.\ Rev.\ D {\bf 16}, 1791 (1977).
  doi:10.1103/PhysRevD.16.1791
  
\bibitem{Weinberg:1977ma} 
  S.~Weinberg,
  Phys.\ Rev.\ Lett.\  {\bf 40}, 223 (1978).
  doi:10.1103/PhysRevLett.40.223

\bibitem{Wilczek:1977pj} 
  F.~Wilczek,
  Phys.\ Rev.\ Lett.\  {\bf 40}, 279 (1978).
  doi:10.1103/PhysRevLett.40.279
  
\bibitem{Bardeen:1986yb} 
  W.~A.~Bardeen, R.~D.~Peccei and T.~Yanagida,
  Nucl.\ Phys.\ B {\bf 279}, 401 (1987).
  doi:10.1016/0550-3213(87)90003-4
  
\bibitem{Kim:1979if} 
  J.~E.~Kim,
  Phys.\ Rev.\ Lett.\  {\bf 43}, 103 (1979).
  doi:10.1103/PhysRevLett.43.103
  
\bibitem{Shifman:1979if} 
  M.~A.~Shifman, A.~I.~Vainshtein and V.~I.~Zakharov,
  Nucl.\ Phys.\ B {\bf 166}, 493 (1980).
  doi:10.1016/0550-3213(80)90209-6
  
\bibitem{Dine:1981rt} 
  M.~Dine, W.~Fischler and M.~Srednicki,
  Phys.\ Lett.\  {\bf 104B}, 199 (1981).
  doi:10.1016/0370-2693(81)90590-6

\bibitem{Preskill:1982cy} 
  J.~Preskill, M.~B.~Wise and F.~Wilczek,
  Phys.\ Lett.\  {\bf 120B}, 127 (1983).
  doi:10.1016/0370-2693(83)90637-8

\bibitem{Abbott:1982af} 
  L.~F.~Abbott and P.~Sikivie,
  Phys.\ Lett.\  {\bf 120B}, 133 (1983).
  doi:10.1016/0370-2693(83)90638-X
  
\bibitem{Dine:1982ah} 
  M.~Dine and W.~Fischler,
  Phys.\ Lett.\  {\bf 120B}, 137 (1983).
  doi:10.1016/0370-2693(83)90639-1
  
\bibitem{Bae:2008ue} 
  K.~J.~Bae, J.~H.~Huh and J.~E.~Kim,
  JCAP {\bf 0809}, 005 (2008)
  doi:10.1088/1475-7516/2008/09/005
  [arXiv:0806.0497 [hep-ph]].
  
  \bibitem{ArkaniHamed:2008qp} 
  N.~Arkani-Hamed and N.~Weiner,
  JHEP {\bf 0812}, 104 (2008)
  doi:10.1088/1126-6708/2008/12/104
  [arXiv:0810.0714 [hep-ph]].

\bibitem{Cheung:2009qd} 
  C.~Cheung, J.~T.~Ruderman, L.~T.~Wang and I.~Yavin,
  Phys.\ Rev.\ D {\bf 80}, 035008 (2009)
  doi:10.1103/PhysRevD.80.035008
  [arXiv:0902.3246 [hep-ph]].

\bibitem{Lee:2016ejx} 
  H.~S.~Lee and M.~S.~Seo,
  Phys.\ Lett.\ B {\bf 767}, 69 (2017)
  doi:10.1016/j.physletb.2017.01.058
  [arXiv:1608.02708 [hep-ph]].

\bibitem{Fayet:1980rr} 
  P.~Fayet,
  Nucl.\ Phys.\ B {\bf 187}, 184 (1981).
  doi:10.1016/0550-3213(81)90122-X
  
  \bibitem{ArkaniHamed:2008qn} 
  N.~Arkani-Hamed, D.~P.~Finkbeiner, T.~R.~Slatyer and N.~Weiner,
  Phys.\ Rev.\ D {\bf 79}, 015014 (2009)
  doi:10.1103/PhysRevD.79.015014
  [arXiv:0810.0713 [hep-ph]].

\bibitem{Tulin:2012wi} 
  S.~Tulin, H.~B.~Yu and K.~M.~Zurek,
  Phys.\ Rev.\ Lett.\  {\bf 110}, no. 11, 111301 (2013)
  doi:10.1103/PhysRevLett.110.111301
  [arXiv:1210.0900 [hep-ph]].

\bibitem{Gninenko:2001hx} 
  S.~N.~Gninenko and N.~V.~Krasnikov,
  Phys.\ Lett.\ B {\bf 513}, 119 (2001)
  doi:10.1016/S0370-2693(01)00693-1
  [hep-ph/0102222].

\bibitem{Fayet:2007ua} 
  P.~Fayet,
  Phys.\ Rev.\ D {\bf 75}, 115017 (2007)
  doi:10.1103/PhysRevD.75.115017
  [hep-ph/0702176 [HEP-PH]].

\bibitem{Pospelov:2008zw} 
  M.~Pospelov,
  Phys.\ Rev.\ D {\bf 80}, 095002 (2009)
  doi:10.1103/PhysRevD.80.095002
  [arXiv:0811.1030 [hep-ph]].
  
  \bibitem{Holdom:1985ag} 
  B.~Holdom,
  Phys.\ Lett.\  {\bf 166B}, 196 (1986).
  doi:10.1016/0370-2693(86)91377-8
  
\bibitem{darkPhotonCoupling}
For a convenient reference, see Ref.~\cite{Lee:2016ief} and references therein.

\bibitem{Lee:2016ief} 
  H.~S.~Lee and S.~Yun,
  Phys.\ Rev.\ D {\bf 93}, no. 11, 115028 (2016)
  doi:10.1103/PhysRevD.93.115028
  [arXiv:1604.01213 [hep-ph]].

\bibitem{Davoudiasl:2012ag} 
  H.~Davoudiasl, H.~S.~Lee and W.~J.~Marciano,
  Phys.\ Rev.\ D {\bf 85}, 115019 (2012)
  doi:10.1103/PhysRevD.85.115019
  [arXiv:1203.2947 [hep-ph]].
  
\bibitem{Lee:2013fda} 
  H.~S.~Lee and M.~Sher,
  Phys.\ Rev.\ D {\bf 87}, no. 11, 115009 (2013)
  doi:10.1103/PhysRevD.87.115009
  [arXiv:1303.6653 [hep-ph]].

\bibitem{Davoudiasl:2014mqa} 
  H.~Davoudiasl, W.~J.~Marciano, R.~Ramos and M.~Sher,
  Phys.\ Rev.\ D {\bf 89}, no. 11, 115008 (2014)
  doi:10.1103/PhysRevD.89.115008
  [arXiv:1401.2164 [hep-ph]].

\bibitem{Kong:2014jwa} 
  K.~Kong, H.~S.~Lee and M.~Park,
  Phys.\ Rev.\ D {\bf 89}, no. 7, 074007 (2014)
  doi:10.1103/PhysRevD.89.074007
  [arXiv:1401.5020 [hep-ph]].

\bibitem{Kim:2014ana} 
  D.~Kim, H.~S.~Lee and M.~Park,
  JHEP {\bf 1503}, 134 (2015)
  doi:10.1007/JHEP03(2015)134
  [arXiv:1411.0668 [hep-ph]].
  
\bibitem{Hall:2009bx} 
  L.~J.~Hall, K.~Jedamzik, J.~March-Russell and S.~M.~West,
  JHEP {\bf 1003}, 080 (2010)
  doi:10.1007/JHEP03(2010)080
  [arXiv:0911.1120 [hep-ph]].

\bibitem{Baur:2015jsy} 
  J.~Baur, N.~Palanque-Delabrouille, C.~Yèche, C.~Magneville and M.~Viel,
  JCAP {\bf 1608}, no. 08, 012 (2016)
  doi:10.1088/1475-7516/2016/08/012
  [arXiv:1512.01981 [astro-ph.CO]].

\bibitem{Masso:2002np} 
  E.~Masso, F.~Rota and G.~Zsembinszki,
  Phys.\ Rev.\ D {\bf 66}, 023004 (2002)
  doi:10.1103/PhysRevD.66.023004
  [hep-ph/0203221].

\bibitem{Graf:2010tv} 
  P.~Graf and F.~D.~Steffen,
  Phys.\ Rev.\ D {\bf 83}, 075011 (2011)
  doi:10.1103/PhysRevD.83.075011
  [arXiv:1008.4528 [hep-ph]].
  
\bibitem{Salvio:2013iaa} 
  A.~Salvio, A.~Strumia and W.~Xue,
  JCAP {\bf 1401}, 011 (2014)
  doi:10.1088/1475-7516/2014/01/011
  [arXiv:1310.6982 [hep-ph]].
  
\bibitem{Bolz:2000fu} 
  M.~Bolz, A.~Brandenburg and W.~Buchmuller,
  Nucl.\ Phys.\ B {\bf 606}, 518 (2001)
  Erratum: [Nucl.\ Phys.\ B {\bf 790}, 336 (2008)]
  doi:10.1016/S0550-3213(01)00132-8, 10.1016/j.nuclphysb.2007.09.020
  [hep-ph/0012052].
  
  \bibitem{Fukugita:1982ep} 
  M.~Fukugita, S.~Watamura and M.~Yoshimura,
  Phys.\ Rev.\ Lett.\  {\bf 48}, 1522 (1982).
  doi:10.1103/PhysRevLett.48.1522
  
\bibitem{Arias:2012az} 
  P.~Arias, D.~Cadamuro, M.~Goodsell, J.~Jaeckel, J.~Redondo and A.~Ringwald,
  JCAP {\bf 1206}, 013 (2012)
  doi:10.1088/1475-7516/2012/06/013
  [arXiv:1201.5902 [hep-ph]].

\bibitem{Pilaftsis:2003gt} 
  A.~Pilaftsis and T.~E.~J.~Underwood,
  Nucl.\ Phys.\ B {\bf 692}, 303 (2004)
  doi:10.1016/j.nuclphysb.2004.05.029
  [hep-ph/0309342].

\bibitem{Kawasaki:2004qu} 
  M.~Kawasaki, K.~Kohri and T.~Moroi,
  Phys.\ Rev.\ D {\bf 71}, 083502 (2005)
  doi:10.1103/PhysRevD.71.083502
  [astro-ph/0408426].

\bibitem{Essig:2013goa} 
  R.~Essig, E.~Kuflik, S.~D.~McDermott, T.~Volansky and K.~M.~Zurek,
  JHEP {\bf 1311}, 193 (2013)
  doi:10.1007/JHEP11(2013)193
  [arXiv:1309.4091 [hep-ph]].

\bibitem{Boyarsky:2008mt} 
  A.~Boyarsky, J.~Lesgourgues, O.~Ruchayskiy and M.~Viel,
  Phys.\ Rev.\ Lett.\  {\bf 102}, 201304 (2009)
  doi:10.1103/PhysRevLett.102.201304
  [arXiv:0812.3256 [hep-ph]].

\bibitem{Harada:2014lma} 
  A.~Harada and A.~Kamada,
  JCAP {\bf 1601}, no. 01, 031 (2016)
  doi:10.1088/1475-7516/2016/01/031
  [arXiv:1412.1592 [astro-ph.CO]].

\bibitem{Kamada:2016vsc} 
  A.~Kamada, K.~T.~Inoue and T.~Takahashi,
  Phys.\ Rev.\ D {\bf 94}, no. 2, 023522 (2016)
  doi:10.1103/PhysRevD.94.023522
  [arXiv:1604.01489 [astro-ph.CO]].

\bibitem{Demidov:2014hka} 
  S.~V.~Demidov and D.~S.~Gorbunov,
  Phys.\ Rev.\ D {\bf 90}, 035014 (2014)
  doi:10.1103/PhysRevD.90.035014
  [arXiv:1404.1339 [hep-ph]].

\bibitem{Asztalos:2009yp} 
  S.~J.~Asztalos {\it et al.} [ADMX Collaboration],
  Phys.\ Rev.\ Lett.\  {\bf 104}, 041301 (2010)
  doi:10.1103/PhysRevLett.104.041301
  [arXiv:0910.5914 [astro-ph.CO]].

\bibitem{CAPP}
  Y.~K.~Semertzidis,
  {\em CAPP: Axions and proton EDM},
  a presentation at Light Dark World International Forum 2016 (July 2016, Daejeon, Korea).

\bibitem{Wagner:2010mi} 
  A.~Wagner {\it et al.} [ADMX Collaboration],
  Phys.\ Rev.\ Lett.\  {\bf 105}, 171801 (2010)
  doi:10.1103/PhysRevLett.105.171801
  [arXiv:1007.3766 [hep-ex]].

\bibitem{Jaeckel:2015jla} 
  J.~Jaeckel and M.~Spannowsky,
  Phys.\ Lett.\ B {\bf 753}, 482 (2016)
  doi:10.1016/j.physletb.2015.12.037
  [arXiv:1509.00476 [hep-ph]].

\bibitem{Alves:2016koo} 
  A.~Alves, A.~G.~Dias and K.~Sinha,
  JHEP {\bf 1608}, 060 (2016)
  doi:10.1007/JHEP08(2016)060
  [arXiv:1606.06375 [hep-ph]].

\end{thebibliography}
\end{document}